\documentclass[11pt]{article}%
\usepackage{amsfonts}
\usepackage{amssymb}
\usepackage{graphicx}
\usepackage{amsmath}%
\providecommand{\U}[1]{\protect\rule{.1in}{.1in}}
\begin{document}

\title{Bose-Einstein condensation in two-dimensional traps}
\author{Mi Xie\thanks{Email: xiemi@tju.edu.cn}\\{\footnotesize Department of Physics, School of Science, Tianjin University,
Tianjin 300072, P. R. China}}
\date{}
\maketitle

\begin{abstract}
In two-dimensional traps, since the theoretical study of Bose-Einstein
condensation (BEC) will encounter the problem of divergence, the actual
contribution of the divergent terms is often estimated in some indirect ways
with the accuracy to the leading order. In this paper, by using an analytical
continuation method to solve the divergence problem, we obtain the analytical
expressions of critical temperature and condensate fraction for Bose gases in
a two-dimensional anisotropic box and harmonic trap, respectively. They are
consistent with or better than previous studies. Then, we further consider the
nonvanishing chemical potential, and obtain the expressions of chemical
potential and more precise condensate fraction. These results agree with the
numerical calculation well, especially for the case of harmonic traps. The
comparison between the grand canonical and canonical ensembles shows that our
calculation in the grand canonical ensemble is reliable.

\end{abstract}

\section{Introduction}

In recent years, BEC in two-dimensional systems attracts much research. First,
the BEC of cold atoms in (quasi)two-dimensional traps has been realized in
experiments \cite{GVL,KHD,CCB}. Then, more interestingly, the BEC of various
bosonic quasiparticles in many-body systems has been widely investigated, such
as excitons \cite{EM}, magnons \cite{NOO,DDD,GRT}, cavity photons
\cite{KSV,KVW,SDD}, and exciton-polaritons \cite{KRK,BHS,SWY}. Many
experiments of quasiparticles are realized in two-dimensional traps.

In two dimensions, the realization of BEC is mainly in a box or harmonic trap.
In the thermodynamic limit, these two cases have a remarkable difference: As
the temperature descends, an ideal Bose gas in a two-dimensional harmonic trap
will undergo the BEC phase transition, but in two-dimensional infinite space
there is no phase transition. In finite systems, however, their difference
becomes small since genuine phase transition cannot occur in either case. In
both cases, at low enough temperature, a large fraction of particles will fall
into the ground state, so the condensation can still occur. This kind of
condensation phenomenon can be observed in experiments.

Unfortunately, there is an obstacle in the theoretical interpretation of the
influence of trapping potentials or boundaries on the critical temperature of
BEC for ideal Bose gases (We will still use the word 'critical temperature' in
this paper though there is no genuine phase transition in a finite system). In
the thermodynamic limit, the critical temperature is determined by the
condition that the excited-state population $N_{e}$ is equal to the total
particle number $N$ when the chemical potential $\mu=0$. In a finite system,
this condition can still be used as an approximate method. However, for
trapped gases, the expression of $N_{e}$ is usually divergent at $\mu=0$. This
problem is not too serious for a two-dimensional harmonic trap since the
leading term is convergent. By neglecting all the other divergent terms, one
can obtain the zero-order critical temperature, which is actually the result
in the thermodynamic limit and is widely used in the literature
\cite{HHA,BPK,DGP,BLS,BKL}. In a two-dimensional box, the problem is
particularly serious since all terms of $N_{e}$ are divergent at $\mu=0$. Then
even the zero-order result cannot be obtained. In the literature, the critical
temperature is determined by, for example, setting a given condensate fraction
\cite{JJ} or numerical calculation \cite{LGJJ}. To obtain more precise
results, the finite-size effect has been studies for many years, some
approximate results of critical temperature and condensate fraction are also
presented, often based on the analysis of the nonvanishing ground-state energy
in a finite system and only including the leading correction
\cite{KVD,Mullin,Yuk2016}. A systematic method for studying the influence of
potentials and boundaries is still lacking.

In this paper, we will use an analytical continuation method to deal with the
divergence problem at $\mu=0$, which is based on the heat kernel expansion and
$\zeta$-function regularization \cite{Xie}. First, we will show that the
divergence can be removed by a general treatment, and the analytical
expressions for critical temperature and condensate fraction for ideal Bose
gases in a two-dimensional anisotropic box or harmonic trap are presented,
respectively. These results are consistent with or better than the previous
studies. Then, more precisely, $\mu=0$ does not exactly hold below the
transition point in a finite system, but the divergence problem makes it
difficult to solve the chemical potential. We will show that our method is
applicable to this problem, and we will give the analytic expressions of the
chemical potential and the more precise condensate fraction, respectively.
These results agree with the numerical calculation well, especially for the
harmonic traps. In addition, to check the influence of the fluctuation in the
grand canonical ensemble, we compare the condensate fraction in the grand
canonical and canonical ensembles. The comparison indicates that the
difference between these two ensembles is very small for particle number
$N\sim10^{3}$.

The paper is organized as follows. In section \ref{sec2}, we discuss the BEC
of an ideal Bose gas in a two-dimensional rectangle box. The analytical
expressions of the critical temperature, the condensate fraction, and the
chemical potential are obtained. In section \ref{sec3}, we discuss the Bose
gas in a two-dimensional anisotropic harmonic trap. The first-order correction
to the critical temperature, and the analytical expressions of condensate
fraction and chemical potential are obtained.\ They agree with the numerical
results very well. In section \ref{sec_CE}, we give a comparison between the
grand canonical and canonical ensembles to show the influence of fluctuation
in the grand canonical ensemble. The conclusion and some discussion are
presented in section \ref{sec4}. A kind of the Epstein $\zeta$-function is
used in our calculation, so we give its asymptotic expansion in Appendix A.

\section{Two-dimensional rectangle box\label{sec2}}

The main tool used in this paper is the heat kernel expansion. In the grand
canonical ensemble, the average particle number of an ideal Bose gas can be
expanded as%
\begin{equation}
\left\langle N\right\rangle =\sum_{i}\frac{1}{z^{-1}e^{\beta E_{i}}-1}%
=\sum_{\ell=1}^{\infty}z^{\ell}\sum_{i}e^{-\ell\beta E_{i}}=\sum_{\ell
=1}^{\infty}K\left(  \ell\frac{\hbar^{2}}{2m}\beta\right)  z^{\ell
},\label{N_box}%
\end{equation}
where $z=e^{\beta\mu}$ is the fugacity, $\beta=1/\left(  k_{B}T\right)  $ with
$k_{B}$ denoting the Boltzmann constant, $\left\{  E_{i}\right\}  $ is the
single-particle energy spectrum, which is proportional to the spectrum
$\left\{  \lambda_{i}\right\}  $ of the Laplacian operator $D=-\nabla
^{2}+\left(  2m/\hbar^{2}\right)  V\left(  \mathbf{x}\right)  $,
$E_{i}=\left(  \hbar^{2}/2m\right)  \lambda_{i}$, and $K\left(  t\right)  $
denotes the global heat kernel of the operator $D$
\cite{KirstenBK,Vassilevich,Gilkey}%
\begin{equation}
K\left(  t\right)  =\sum_{i=0}^{\infty}e^{-\lambda_{i}t}.\label{Kt}%
\end{equation}
For small $t$, the heat kernel expansion of $K\left(  t\right)  $ has the
asymptotic form \cite{KirstenBK,Vassilevich,Gilkey}%
\begin{equation}
K\left(  t\right)  \approx\frac{1}{\left(  4\pi t\right)  ^{d/2}}%
\sum_{k=0,\frac{1}{2},1,\cdots}^{\infty}B_{k}t^{k},\text{ \ \ \ }\left(
t\rightarrow0\right)
\end{equation}
where $d$ is the spatial dimension and $B_{k}$ $\left(  k=0,1/2,1,\cdots
\right)  $ are the heat kernel coefficients. Thus, eq. (\ref{N_box}) expresses
the average particle number of the Bose gas as a series of global heat kernels.

In the thermodynamic limit, the critical temperature of BEC is determined by
the condition that the excited-state population $N_{e}$ equals the total
particle number $N$ at $\mu=0$. In a finite system, although genuine phase
transitions cannot occur, we can expect to obtain the critical temperature by
the same condition as an approximation.

The excited-state population is easy to find from eq. (\ref{N_box}) by
excluding the ground-state contribution. Furthermore, the transition occurring
at $\mu=0$ means that the ground-state energy should be zero, so we need to
shift the energy spectrum so that the ground-state energy vanishes. In other
words, we will replace the heat kernel eq. (\ref{Kt}) by
\begin{equation}
K^{\prime}\left(  t\right)  =\sum_{i=1}^{\infty}e^{-\left(  \lambda
_{i}-\lambda_{0}\right)  t},\label{Kpt}%
\end{equation}
in which the ground-state contribution is excluded. Therefore, for the
two-dimensional case, the corresponding heat kernel coefficients change to%
\begin{equation}
B_{0}^{\prime}=B_{0},B_{1/2}^{\prime}=B_{1/2},B_{1}^{\prime}=B_{1}+\lambda
_{0}B_{0}-4\pi,\cdots\label{Bkp2D}%
\end{equation}

In the following, we will consider a Bose gas in a two-dimensional rectangle
box of length sides $L_{x}$ and $L_{y}$ with Dirichlet boundary conditions.
The shifted spectrum is%
\begin{equation}
\lambda\left(  n_{x},n_{y}\right)  =\pi^{2}\left(  \frac{n_{x}^{2}}{L_{x}^{2}%
}+\frac{n_{y}^{2}}{L_{y}^{2}}\right)  -\pi^{2}\left(  \frac{1}{L_{x}^{2}%
}+\frac{1}{L_{y}^{2}}\right)  .\text{ \ \ \ \ }\left(  n_{x},n_{y}%
=1,2,3,\cdots\right)
\end{equation}
According to eq. (\ref{Bkp2D}) and the usual heat kernel coefficients
\cite{DX2009}, the heat kernel coefficients for $K^{\prime}\left(  t\right)  $
are%
\begin{equation}
B_{0}^{\prime}=S=L_{x}L_{y},B_{1/2}^{\prime}=-\sqrt{\pi}\left(  L_{x}%
+L_{y}\right)  ,B_{1}^{\prime}=\pi^{2}\left(  \frac{L_{y}}{L_{x}}+\frac{L_{x}%
}{L_{y}}\right)  -3\pi,\cdots.\label{Bkp}%
\end{equation}
Replacing the $K\left(  t\right)  $ in eq. (\ref{N_box}) by $K^{\prime}\left(
t\right)  $, we can obtain the excited-state population as%
\begin{equation}
N_{e}=\sum_{\ell=1}^{\infty}K^{\prime}\left(  \ell\frac{\hbar^{2}}{2m}%
\beta\right)  z^{\ell}=\sum_{k=0,\frac{1}{2},1,\cdots}^{\infty}\frac
{B_{k}^{\prime}}{\left(  4\pi\right)  ^{k}}\lambda^{2k-2}g_{1-k}\left(
z\right)  ,\label{Nexp}%
\end{equation}
where%
\begin{equation}
g_{\sigma}\left(  z\right)  =\frac{1}{\Gamma(\sigma)}\int_{0}^{\infty}%
\frac{x^{\sigma-1}}{z^{-1}e^{x}-1}dx=\sum_{k=1}^{\infty}\frac{z^{k}}%
{k^{\sigma}}\label{BEexp}%
\end{equation}
is the Bose-Einstein integral, and $\lambda=\sqrt{2\pi\beta}\hbar/\sqrt{m}$ is
the mean thermal wavelength. In eq. (\ref{Nexp}) we have replaced
$\left\langle N_{e}\right\rangle $ by $N_{e}$ for simplicity.

In eq. (\ref{Nexp}), the heat kernel coefficient $B_{k}^{\prime}$ has a
dimension of $\left[  L^{2-2k}\right]  $. If we denote the characteristic
length scale of the system as $\bar{L}$, $B_{k}^{\prime}$ will be roughly
proportional to $\bar{L}^{2-2k}$, just as in eq. (\ref{Bkp}). Therefore, eq.
(\ref{Nexp}) is in fact a series of $\lambda/\bar{L}$.

\subsection{Critical temperature $T_{c}$}

The critical temperature of BEC is determined by $N_{e}=N$ at $\mu=0$. In eq.
(\ref{Nexp}), $N_{e}$ is expressed as a series of a small parameter
$\lambda/\bar{L}$, so usually the higher-order terms are just small
corrections. However, when $\mu\rightarrow0$, since the asymptotic behavior of
the Bose-Einstein integral is
\begin{equation}
g_{\sigma}\left(  e^{\beta\mu}\right)  \approx\left\{
\begin{array}
[c]{lll}%
\zeta\left(  \sigma\right)  , & \left(  \sigma\geq\frac{3}{2}\right)   & \\
-\ln\left(  -\beta\mu\right)  , & \left(  \sigma=1\right)   & \\
\Gamma\left(  1-\sigma\right)  \frac{1}{\left(  -\beta\mu\right)  ^{1-\sigma}%
},\text{ } & \left(  \sigma\leq\frac{1}{2}\right)   & \text{ \ \ \ }\left(
\mu\rightarrow0\right)
\end{array}
\right.  \label{BEint}%
\end{equation}
where $\zeta\left(  \sigma\right)  =\sum_{n=1}^{\infty}n^{-\sigma}$ is the
Riemann zeta function, every term in eq. (\ref{Nexp}) is divergent, and the
divergence becomes more severe in the higher orders. As a result, it will not
work to truncate this series at any finite order. To overcome this divergence
problem, we will use an analytical continuation method with the help of the
heat kernel expansion and $\zeta$-function regularization \cite{Xie}, in which
all the terms in the series are considered.

First, substituting the leading term in the asymptotic expansion of the
Bose-Einstein integral eq. (\ref{BEint}) into eq. (\ref{Nexp}) gives
\begin{equation}
n_{e}\lambda^{2}\approx-\ln\left(  -\beta\mu\right)  +\sum_{k=\frac{1}%
{2},1,\cdots}^{\infty}\Gamma\left(  k\right)  \frac{B_{k}^{\prime}}{\left(
4\pi\right)  ^{k}S}\lambda^{2k}\frac{1}{\left(  -\beta\mu\right)  ^{k}},
\label{nlambda0}%
\end{equation}
where $n_{e}=N_{e}/S$ is the number density of excited-state particles. We
hope to express the divergent sum in eq. (\ref{nlambda0}) by the heat kernel.
For this purpose, introduce a regularization parameter $s$ which will be set
to $0$ at the end of the calculation in the gamma function%
\begin{equation}
\Gamma\left(  \xi\right)  =\int_{0}^{\infty}x^{\xi-1+s}e^{-x}dx.\text{
\ \ \ }\left(  s\rightarrow0\right)  \label{Gamma}%
\end{equation}
Eq. (\ref{nlambda0}) becomes%
\begin{align}
n_{e}\lambda^{2}  &  =-\ln\left(  -\beta\mu\right)  +\int_{0}^{\infty
}dxx^{-1+s}e^{-x}\left[  \frac{1}{S}\sum_{k=0,\frac{1}{2},1,\cdots}^{\infty
}B_{k}^{\prime}\left(  \frac{\hbar^{2}}{2m\left(  -\mu\right)  }x\right)
^{k}-1\right] \nonumber\\
&  =-\ln\left(  -\beta\mu\right)  +\frac{2\pi\hbar^{2}}{mS\left(  -\mu\right)
}\int_{0}^{\infty}dxx^{s}e^{-x}K^{\prime}\left(  \frac{\hbar^{2}}{2m\left(
-\mu\right)  }x\right)  -\Gamma\left(  s\right)  . \label{ac}%
\end{align}
In the last line we have replaced the divergent series by the heat kernel
$K^{\prime}\left(  t\right)  $ according to the heat kernel expansion.

Then, by the definition of heat kernel eq. (\ref{Kpt}), we can perform the
integral in eq. (\ref{ac}),
\begin{equation}
n_{e}\lambda^{2}=-\ln\left(  -\beta\mu\right)  +\frac{2\pi\hbar^{2}}{mS}%
\Gamma\left(  1+s\right)  \left(  -\mu\right)  ^{s}\sum\nolimits^{\prime}%
\frac{1}{\left[  E\left(  n_{x},n_{y}\right)  -\mu\right]  ^{1+s}}%
-\Gamma\left(  s\right)  ,\label{nlambda}%
\end{equation}
where the prime on the sum $\sum\nolimits^{\prime}$ denotes that the ground
state is excluded. Since the transition occurs at $\mu=0$, by neglecting the
chemical potential $\mu$ in the denominator, the sum in eq. (\ref{nlambda})
becomes%
\begin{align}
&  \sum\nolimits^{\prime}\frac{1}{\left[  E\left(  n_{x},n_{y}\right)
\right]  ^{1+s}}=\left(  \frac{2mS}{\pi^{2}\hbar^{2}}\right)  ^{1+s}%
\sum_{\left(  n_{x},n_{y}\right)  \neq\left(  1,1\right)  }^{\infty}\frac
{1}{\left[  \left(  \chi^{-1}n_{x}^{2}+\chi n_{y}^{2}\right)  -\left(
\chi+\chi^{-1}\right)  \right]  ^{1+s}}\nonumber\\
&  =\left(  \frac{2mS}{\pi^{2}\hbar^{2}}\right)  ^{1+s}\sum_{p=0}^{\infty
}\left(
\begin{array}
[c]{c}%
p+s\\
p
\end{array}
\right)  \left[  \left(  \chi+\chi^{-1}\right)  ^{p}E_{2}\left(
1+s+p;\chi^{-1},\chi\right)  -\left(  \chi+\chi^{-1}\right)  ^{-1-s}\right]
,\label{sum0}%
\end{align}
where we have introduced a shape factor $\chi=L_{x}/L_{y}$, and $\left(
\begin{array}
[c]{c}%
n\\
k
\end{array}
\right)  =\frac{n!}{k!\left(  n-k\right)  !}$ is the binomial coefficient,
\begin{equation}
E_{2}\left(  \sigma;a_{1},a_{2}\right)  =\sum_{n_{1},n_{2}=1}^{\infty}\frac
{1}{\left(  a_{1}n_{1}^{2}+a_{2}n_{2}^{2}\right)  ^{\sigma}}%
\end{equation}
is the Epstein $\zeta$-function. By use of eq. (\ref{E.020}) in Appendix A,
when $s\rightarrow0$, eq. (\ref{sum0}) is divergent and its asymptotic form is%
\begin{equation}
\sum\nolimits^{\prime}\frac{1}{\left[  E\left(  n_{x},n_{y}\right)  \right]
^{1+s}}\approx\left(  \frac{2mS}{\pi^{2}\hbar^{2}}\right)  ^{1+s}\frac{\pi}%
{4}\left(  \frac{1}{s}+\Omega_{2}\right)  ,\label{sum1}%
\end{equation}
where
\begin{align}
\Omega_{2} &  =3\gamma+\psi\left(  \frac{1}{2}\right)  -\frac{\pi}{3}\left(
\chi+\chi^{-1}\right)  -\frac{4}{\pi}\left(  \chi+\chi^{-1}\right)  ^{-1}%
-\ln\left[  \chi\eta^{4}\left(  i\chi\right)  \right]  \nonumber\\
&  +\frac{4}{\pi}\sum_{p=1}^{\infty}\left[  \left(  \chi+\chi^{-1}\right)
^{p}E_{2}\left(  1+p;\chi^{-1},\chi\right)  -\left(  \chi+\chi^{-1}\right)
^{-1}\right]
\end{align}
is a parameter only related to the shape factor $\chi$, $\gamma\approx0.5772$
is the Euler constant, $\psi\left(  z\right)  =\Gamma^{\prime}\left(
z\right)  /\Gamma\left(  z\right)  $ is the digamma function, and
\begin{equation}
\eta\left(  \tau\right)  =e^{\frac{\pi i\tau}{12}}\prod\limits_{n=1}^{\infty
}\left(  1-e^{2n\pi i\tau}\right)
\end{equation}
is the Dedekind $\eta$-function. Since for $s\rightarrow0$,
\begin{equation}
\Gamma\left(  s-n\right)  \approx\frac{\left(  -1\right)  ^{n}}{n!}\left[
\frac{1}{s}+\psi\left(  n+1\right)  \right]  ,\text{ \ \ }\left(
n=0,1,2,\cdots\right)
\end{equation}
the divergent term of $s$ from eq. (\ref{sum1}) and that from the term with
$\Gamma\left(  s\right)  $ are exactly canceled.

Finally, eq. (\ref{nlambda}) becomes%
\begin{equation}
n_{e}\lambda^{2}=\ln\frac{2mS}{\pi^{2}\hbar^{2}\beta}+\Omega_{2}=\ln\frac
{N}{n\lambda^{2}}+\Omega,\label{nlambda-2}%
\end{equation}
where we have introduced
\begin{equation}
\Omega=\ln\frac{4}{\pi}+\Omega_{2}%
\end{equation}
for simplicity. In eq. (\ref{nlambda-2}), all of the divergent terms of $\mu$
are also canceled, and the final result is fully analytical, so the critical
temperature is
\begin{equation}
T_{c}=\frac{2\pi\hbar^{2}}{mk_{B}}\frac{n}{W\left(  Ne^{\Omega}\right)
},\label{Tc_box}%
\end{equation}
where $W\left(  z\right)  $ is the Lambert $W$ function, satisfying
$z=W\left(  ze^{z}\right)  $.

\begin{figure}[ptb]
\begin{center}
\includegraphics[height=3in]{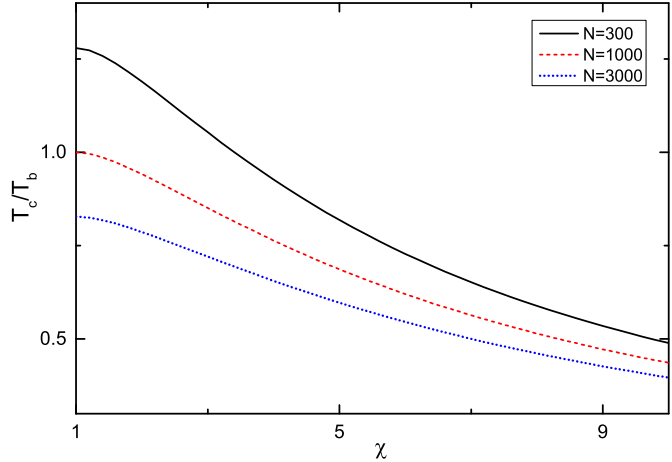}
\end{center}
\caption{The influence of anisotropy on the critical temperature at fixed
density for different $N$ in a two-dimensional box. It shows that the
anisotropy reduces the critical temperature. }%
\label{fig1}%
\end{figure}

Eq. (\ref{Tc_box}) gives the influence of the particle number and the shape of
box on the critical temperature. In fig. \ref{fig1} we plot the relation
between critical temperature and $\chi$ at fixed density of particles. It
shows that the anisotropy lowers the critical temperature. In this and the
following figures, the temperature is rescaled to $T/T_{b}$, where
\begin{equation}
T_{b}=T_{c}\left(  N=1000,\chi=1\right)  \approx0.228\frac{2\pi\hbar^{2}%
n}{mk_{B}}%
\end{equation}
is the critical temperature for $N=1000$ in a square box.

\begin{figure}[ptb]
\begin{center}
\includegraphics[height=3in]{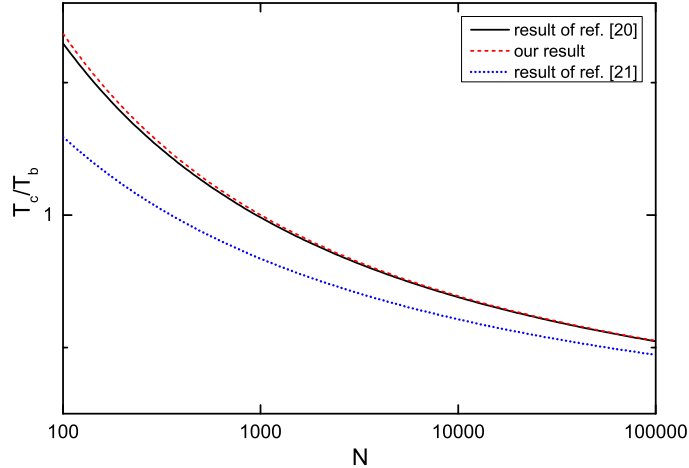}
\end{center}
\caption{The relation between critical temperature and particle number at
fixed density in a two-dimensional square box ($\chi=1$). Our result agrees
with the numerical solution in \cite{LGJJ} very well.}%
\label{fig0}%
\end{figure}

There are many studies on the BEC in cavities, most of them concentrate on the
three-dimensional cases \cite{GH,KT}. For two-dimensional boxes, in
\cite{KVD}, the authors give a relation between the critical temperature and
particle number, which is similar to eq. (\ref{nlambda-2}) but with $\Omega
=0$. In ref. \cite{LGJJ}, the authors discuss the property of an ideal Bose
gas in a square box in both the grand canonical ensemble and canonical
ensemble in details. Their research is based on numerical calculation, and
obtain an expression of critical temperature by fitting the numerical
solution. By taking $\chi=1$ so that $\Omega=-1.0468$ in eq. (\ref{Tc_box}),
our result will go back to the square box case. The relation between the
critical temperature and particle number given by eq. (\ref{Tc_box}) and refs.
\cite{LGJJ} and \cite{KVD} are shown in fig. \ref{fig0}. Our result agrees
with the numerical calculation in ref. \cite{LGJJ} quite well.

\subsection{Condensate fraction and chemical potential}

In the above discussion, the chemical potential $\mu$ is assumed to be zero at
the transition point. It implies that $\mu=0$ holds for $T<T_{c}$ just like in
the thermodynamic limit case. Under this assumption, the condensate fraction
can be directly obtained from eq. (\ref{nlambda-2}):%
\begin{equation}
\frac{N_{0}^{\left(  0\right)  }}{N}=1-\frac{1}{n\lambda^{2}}\left(  \ln
\frac{N}{n\lambda^{2}}+\Omega\right)  ,
\end{equation}
which will be called the zero-order condensate fraction in this paper.

The chemical potential $\mu$ cannot be exactly zero at $T<T_{c}$ in a finite
system, but because of the divergence problem, directly solving $\mu$ is
difficult, especially near the transition point. When $T\ll T_{c}$, $\mu$ can
be approximate to $-k_{B}T/N_{0}^{\left(  0\right)  }$, but this approximation
is invalid for $T\sim T_{c}$ since $N_{0}^{\left(  0\right)  }=0$ at the
transition point.

The discussion in the above section provides a way to avoid the divergence, so
we can solve $\mu$ by the similar way. Specifically, accurate to $\mu^{1}$, we
will add three more terms in eq. (\ref{nlambda}) to obtain the expression of
total particle number: the contribution from the ground-state particles%
\begin{equation}
N_{0}^{\left(  1\right)  }=\frac{1}{e^{-\beta\mu}-1}\approx\frac{1}{-\beta\mu
},\label{N0}%
\end{equation}
the next-to-leading term in the asymptotic expansion of the Bose-Einstein
integral in the first term%
\begin{equation}
g_{1}\left(  e^{\beta\mu}\right)  \approx-\ln\left(  -\beta\mu\right)
+\frac{-\beta\mu}{2},\label{g1}%
\end{equation}
and the first-order contribution of $\mu$ in the sum of energy spectrum%
\begin{equation}
\sum\nolimits^{\prime}\frac{1}{\left[  E\left(  n_{x},n_{y}\right)
-\mu\right]  ^{1+s}}\approx\sum\nolimits^{\prime}\left\{  \frac{1}{\left[
E\left(  n_{x},n_{y}\right)  \right]  ^{1+s}}-\frac{\left(  1+s\right)
\left(  -\mu\right)  }{\left[  E\left(  n_{x},n_{y}\right)  \right]  ^{2+s}%
}\right\}  .\label{zeta1_box}%
\end{equation}
In the right-hand side of this equation, the first sum has been given in eq.
(\ref{sum1}); the second sum is analytical at $s=0$, so we can directly set
$s=0$ in it. By introducing a parameter only related to $\chi$,%
\begin{equation}
\Omega_{3}=\sum_{p=0}^{\infty}\left(  p+1\right)  \left(  \chi+\chi
^{-1}\right)  ^{p}\left[  E_{2}\left(  p+2;\chi^{-1},\chi\right)  -\left(
\chi+\chi^{-1}\right)  ^{-p-2}\right]  ,
\end{equation}
we can express the asymptotic expansion of eq. (\ref{zeta1_box}) at
$s\rightarrow0$ as
\begin{equation}
\sum\nolimits^{\prime}\frac{1}{\left[  E\left(  n_{x},n_{y}\right)
-\mu\right]  ^{1+s}}\approx\left(  \frac{2mS}{\pi^{2}\hbar^{2}}\right)
^{1+s}\left[  \frac{\pi}{4}\frac{1}{s}+\frac{\pi}{4}\Omega_{2}-\frac
{2mS\left(  -\mu\right)  }{\pi^{2}\hbar^{2}}\Omega_{3}\right]  .
\end{equation}
Thus, eq. (\ref{nlambda}) with the additional terms becomes%
\begin{equation}
n\lambda^{2}\approx\frac{n\lambda^{2}}{N}\frac{1}{-\beta\mu}-\left(
\frac{16\Omega_{3}}{\pi^{2}}\frac{N}{n\lambda^{2}}-\frac{1}{2}\right)  \left(
-\beta\mu\right)  +\ln\frac{N}{n\lambda^{2}}+\Omega,
\end{equation}
where the divergent terms of $s$ have also been canceled. The term $-1/2$ in
the parentheses in the second term can be neglected, which means that the
contribution from the second term of $g_{1}\left(  e^{\beta\mu}\right)  $ in
eq. (\ref{g1}) is much smaller than that from the second term in the
right-hand side of eq. (\ref{zeta1_box}). After neglecting this small term, we
can solve the chemical potential as%
\begin{equation}
\mu\approx\mu_{c}\left[  \sqrt{1+\frac{\pi^{2}}{64\Omega_{3}}\left(
n\lambda^{2}\right)  ^{2}\left(  \frac{N_{0}^{\left(  0\right)  }}{N}\right)
^{2}}-\frac{\pi}{8\sqrt{\Omega_{3}}}n\lambda^{2}\frac{N_{0}^{\left(  0\right)
}}{N}\right]  ,\label{mu-box}%
\end{equation}
where
\begin{equation}
\mu_{c}=-\frac{\pi^{2}\hbar^{2}}{2\sqrt{\Omega_{3}}mS}%
\end{equation}
is the chemical potential at the transition point.

\begin{figure}[ptb]
\begin{center}
\includegraphics[height=3in]{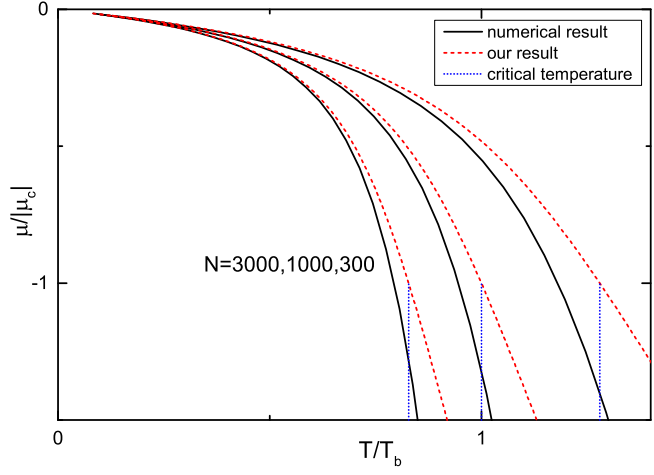}
\end{center}
\caption{The relation between chemical potential and temperature for different
$N$ in a two-dimensional square box ($\chi=1$). The three sets of lines denote
$N=3000,1000,300$ from left to right.}%
\label{fig2}%
\end{figure}

In fig. \ref{fig2} we plot the relation between the chemical potential and
temperature given by eq. (\ref{mu-box}) for different $N$. The result for
$\mu$ in the literature is rare, and we include the numerical results in the
figure for comparison.

\begin{figure}[ptb]
\begin{center}
\includegraphics[height=3in]{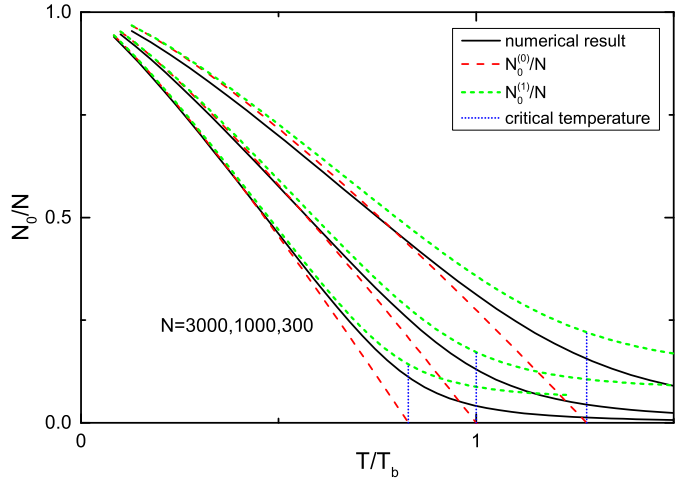}
\end{center}
\caption{The relation between condensate fraction and temperature for
different $N$ in a two-dimensional square box ($\chi=1$). The numerical
solution, the zero- and first-order approximations are plotted for
$N=3000,1000,300$ from left to right. The zero-order condensate fraction
vanishes at $T_{c}$ as excepted.}%
\label{fig3}%
\end{figure}

The first-order condensate fraction $N_{0}^{\left(  1\right)  }/N$ in eq.
(\ref{N0}) is straightforward from eq. (\ref{mu-box}). In fig. \ref{fig3} we
show the relation between the condensate fraction and temperature for
different $N $. We can find that the zero-order condensate fraction
$N_{0}^{\left(  0\right)  }/N$ vanishes at the transition point as expected.

\section{Two-dimensional anisotropic harmonic trap\label{sec3}}

The harmonic trap is the most commonly used trap in BEC experiments and also
in the theoretical research. In fact, the thermodynamic properties of Bose
gases in two-dimensional harmonic traps can be exactly obtained
\cite{Cheng1,Cheng2}. On the other hand, due to the divergence problem at the
transition point, the critical temperature of BEC in a two-dimensional
harmonic trap is often approximately regarded as the thermodynamic-limit value
\cite{HHA,BPK,DGP,BLS,BKL}. In the following we will remove the divergence,
and give the analytical forms of the critical temperature, the condensate
fraction, and the chemical potential.

Consider an ideal Bose gas trapped in an anisotropic harmonic potential%
\begin{equation}
V=\frac{1}{2}m\left(  \omega_{x}^{2}x^{2}+\omega_{y}^{2}y^{2}\right)  .
\end{equation}
The single-particle energy spectrum has the form%
\begin{equation}
E\left(  n_{x},x_{y}\right)  =\hbar\omega_{0}\lambda\left(  n_{x}%
,x_{y}\right)  , \label{E_HO}%
\end{equation}
where $\omega_{0}=\sqrt{\omega_{x}\omega_{y}}$ and%
\begin{equation}
\lambda\left(  n_{x},x_{y}\right)  =\sqrt{\kappa}n_{x}+\frac{1}{\sqrt{\kappa}%
}n_{y},\text{ \ \ \ }\left(  n_{x},n_{y}=0,1,2,\cdots\right)
\end{equation}
where we have introduced $\kappa=\omega_{x}/\omega_{y}$ for convenience, and
the ground-state energy has been shifted to $0$. Consequently, the exact
solution and the asymptotic expansion of the global heat kernel are%
\begin{align}
K\left(  t\right)   &  =\sum\nolimits^{\prime}e^{-\lambda\left(  n_{x}%
,x_{y}\right)  t}=\frac{1}{\left(  1-e^{-\sqrt{\kappa}t}\right)  \left(
1-e^{-t/\sqrt{\kappa}}\right)  }-1\nonumber\\
&  =\sum_{k=0}^{\infty}C_{k}t^{k-2},\text{ \ \ \ \ \ \ }\left(  t\rightarrow
0\right)  \label{Kt_HO}%
\end{align}
where $\sum\nolimits^{\prime}$ still represents that the ground state is
excluded in the sum, and the expansion coefficients are%
\begin{equation}
C_{0}=1,C_{1}=\frac{1}{2}\left(  \sqrt{\kappa}+\frac{1}{\sqrt{\kappa}}\right)
,C_{2}=\frac{1}{12}\left(  \kappa+\frac{1}{\kappa}\right)  -\frac{3}{4}%
,\cdots.
\end{equation}
In such a trap, the excited-state population of an ideal Bose gas is%
\begin{equation}
N_{e}=\sum_{\ell=1}^{\infty}z^{\ell}K\left(  \ell\beta\hbar\omega_{0}\right)
=\sum_{k=0}^{\infty}C_{k}\left(  \beta\hbar\omega_{0}\right)  ^{k-2}%
g_{2-k}\left(  z\right)  . \label{Ne_HO}%
\end{equation}

\subsection{Critical temperature $T_{c}$}

To determine the critical temperature, we need to know the value of eq.
(\ref{Ne_HO}) at $\mu=0$. However, under this condition, except the first term
of eq. (\ref{Ne_HO}), all the other ones are divergent. This divergence can
also be removed by the method used in last section.

First, substituting the leading term of the asymptotic expansion of the
Bose-Einstein integral eq. (\ref{BEint}) into eq. (\ref{Ne_HO}) and replacing
the gamma function by eq. (\ref{Gamma}), we have
\begin{equation}
N_{e}\approx\frac{C_{0}}{\left(  \beta\hbar\omega_{0}\right)  ^{2}}%
\zeta\left(  2\right)  -\frac{C_{1}}{\beta\hbar\omega_{0}}\ln\left(  -\beta
\mu\right)  +I_{2},\label{N_HO}%
\end{equation}
where
\begin{equation}
I_{2}=\frac{1}{\left(  -\beta\mu\right)  }\int_{0}^{\infty}dxx^{s}%
e^{-x}K\left(  \frac{\hbar\omega_{0}}{-\mu}x\right)  -\frac{C_{0}\Gamma\left(
s-1\right)  }{\left(  \beta\hbar\omega_{0}\right)  ^{2}}\left(  -\beta
\mu\right)  -\frac{C_{1}\Gamma\left(  s\right)  }{\beta\hbar\omega_{0}%
}.\label{I2_HO}%
\end{equation}
The integral in the first term becomes a sum over the spectrum,%
\begin{equation}
\frac{1}{\left(  -\beta\mu\right)  }\int_{0}^{\infty}dxx^{s}e^{-x}K\left(
\frac{\hbar\omega_{0}}{-\mu}x\right)  =\frac{\Gamma\left(  1+s\right)  \left(
-\mu\right)  ^{s}}{\beta\left(  \hbar\omega_{y}\right)  ^{1+s}}\sum
\nolimits^{\prime}\frac{1}{\left(  \kappa n_{x}+n_{y}-\frac{\mu}{\hbar
\omega_{y}}\right)  ^{1+s}}.\label{sum2}%
\end{equation}
For simplicity, we assume that $\kappa$ is an integer. For $\mu=0$, the sum
then becomes%
\begin{equation}
\sum\nolimits^{\prime}\frac{1}{\left(  \kappa n_{x}+n_{y}\right)  ^{1+s}}%
=\sum_{n=1}^{\infty}\sum_{n_{x}=0}^{\left[  \frac{n}{\kappa}\right]  }\frac
{1}{n^{1+s}},
\end{equation}
where $n=\kappa n_{x}+n_{y}$, and $\left[  x\right]  $ denotes the greatest
integer not exceeding $x$. Thus,%
\begin{align}
&  \sum\nolimits^{\prime}\frac{1}{\left(  \kappa n_{x}+n_{y}\right)  ^{1+s}%
}\nonumber\\
&  =\sum_{n=1}^{\infty}\left(  \frac{n}{\kappa}+1\right)  \frac{1}{n^{1+s}%
}-\sum_{k=1}^{\kappa-1}\frac{k}{\kappa}\sum_{p=0}^{\infty}\frac{1}{\left(
p\kappa+k\right)  ^{1+s}}\nonumber\\
&  =\frac{1}{\kappa}\zeta\left(  s\right)  +\zeta\left(  1+s\right)
-\sum_{k=1}^{\kappa-1}\frac{k}{\kappa^{2+s}}\zeta\left(  1+s,\frac{k}{\kappa
}\right)  ,\label{sum3}%
\end{align}
where%
\begin{equation}
\zeta\left(  s,a\right)  =\sum_{n=0}^{\infty}\frac{1}{\left(  n+a\right)
^{s}},\left(  a\neq0,-1,-2,\cdots\right)
\end{equation}
is the Hurwitz $\zeta$-function. Eq. (\ref{sum3}) is divergent at
$s\rightarrow0$, but the divergent term is exactly canceled by another
divergent term coming from $\Gamma\left(  s\right)  $ in the last term in eq.
(\ref{I2_HO}). Asymptotically expanding eq. (\ref{I2_HO}) at $s\rightarrow0$
and dropping the term proportional to $\mu$, we have%
\begin{equation}
I_{2}=\frac{1}{2\beta\hbar\omega_{y}}\left[  \left(  1+\frac{1}{\kappa
}\right)  \ln\frac{-\mu}{\hbar\omega_{y}}-\Delta\right]  ,\label{I2_1}%
\end{equation}
where%
\begin{equation}
\Delta=\frac{1}{2\kappa}-\frac{1}{2}\left(  1-\frac{1}{\kappa}\right)
\ln\kappa-\sum_{k=1}^{\kappa-1}\frac{k}{\kappa^{2}}\psi\left(  \frac{k}%
{\kappa}\right)  -\gamma
\end{equation}
is a parameter only related to $\kappa$. Then eq. (\ref{N_HO}) becomes%
\begin{equation}
N_{e}=\frac{\zeta\left(  2\right)  }{\kappa\left(  \beta\hbar\omega
_{y}\right)  ^{2}}+\frac{1}{2\beta\hbar\omega_{y}}\left[  \left(  1+\frac
{1}{\kappa}\right)  \ln\left(  \frac{1}{\beta\hbar\omega_{y}}\right)
-2\Delta\right]  .\label{N3_HO}%
\end{equation}
In this equation, both of the divergent terms of $s$ and $\mu$ are canceled,
so the critical temperature can be obtained analytically by setting $N_{e}=N$.
Compared with the thermodynamic-limit result, the second term in the
right-hand side in eq. (\ref{N3_HO}) is an extra correction. When the
correction is small, the critical temperature is approximately%
\begin{equation}
T_{c}\approx T_{0}\left\{  1-\frac{\sqrt{6\kappa}}{8\pi}\frac{1}{\sqrt{N}%
}\left[  \left(  1+\frac{1}{\kappa}\right)  \left(  \ln N+\ln\frac{6\kappa
}{\pi^{2}}\right)  -4\Delta\right]  \right\}  ,\label{Tc_HO}%
\end{equation}
where
\begin{equation}
T_{0}=\frac{\sqrt{6N}}{\pi}\frac{\hbar\omega_{0}}{k_{B}}\label{T0_HO}%
\end{equation}
is the critical temperature in the thermodynamic limit. The leading term in
the correction to the critical temperature is proportional to $\ln N/\sqrt{N}%
$, which is consistent with the leading term of the quantum correction given
in ref. \cite{Yuk2016}.

\begin{figure}[ptb]
\begin{center}
\includegraphics[height=3in]{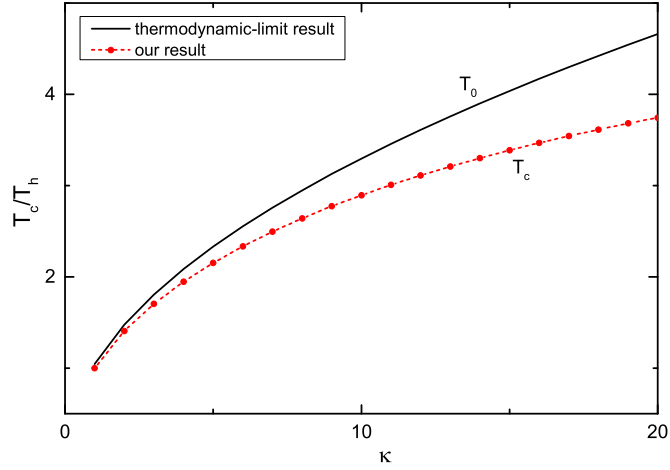}
\end{center}
\caption{The influence of anisotropy on the critical temperature in a
two-dimensional harmonic trap for $N=1000$. The anisotropy increases the
difference between $T_{c}$ and $T_{0}$.}%
\label{fig5}%
\end{figure}

In fig. \ref{fig5}, we plot the critical temperatures eqs. (\ref{Tc_HO}) and
(\ref{T0_HO}) for different $\kappa$. It shows that our result is lower than
the thermodynamic-limit value ($T_{c}<T_{0}$), and the anisotropy increases
the difference between them. In this and the following figures, the
temperature is rescaled to $T/T_{h}$, where $T_{h}=T_{c}\left(  N=1000,\kappa
=1\right)  $.

\subsection{Condensate fraction and chemical potential}

Under the assumption $\mu=0$, the zero-order condensate fraction is easy to
obtain from eq. (\ref{N3_HO}),
\begin{equation}
\frac{N_{0}^{\left(  0\right)  }}{N}=1-\left(  \frac{T}{T_{0}}\right)
^{2}-\frac{\sqrt{6}\left(  \kappa+1\right)  }{4\pi\sqrt{\kappa N}}\frac
{T}{T_{0}}\left\{  \ln\left[  \frac{6\kappa}{\pi^{2}}\left(  \frac{T}{T_{0}%
}\right)  ^{2}N\right]  -\frac{4\kappa}{\kappa+1}\Delta\right\}
.\label{N00_HO}%
\end{equation}
For the isotropic case, i.e. $\kappa=1$, neglecting the higher-order
contribution in the third term, the zero-order condensate fraction can be
expressed as%
\begin{equation}
\frac{N_{0}^{\left(  0\right)  }}{N}\approx1-\left(  \frac{T}{T_{0}}\right)
^{2}-\frac{\sqrt{6}}{2\pi}\frac{T}{T_{0}}\frac{\ln N}{\sqrt{N}}%
.\label{N00iso_HO}%
\end{equation}
In Ref. \cite{Mullin}, the author gives an approximate result of the
condensate fraction in an isotropic harmonic trap, which has the similar form
as eq. (\ref{N00iso_HO}) but the coefficient of the third term is twice as
large as our result. The comparison with the numerical calculation confirms
that eq. (\ref{N00iso_HO}) is much more precise (see fig. \ref{fig7}).

In a finite system, the chemical potential $\mu$ is not exactly zero below the
transition point. To find the analysis form of $\mu$, we need to add three
terms in eq. (\ref{N_HO}) to give an equation of $N$: the ground-state
particle, the next-to-leading term of the Bose-Einstein integral, and the
first-order correction of $\mu$ in eq. (\ref{sum2}). Thus eq. (\ref{N_HO})
becomes%
\begin{align}
N &  \approx\frac{1}{-\beta\mu}+\frac{C_{0}}{\left(  \beta\hbar\omega
_{0}\right)  ^{2}}\left[  \zeta\left(  2\right)  -\beta\mu\left(  \ln\left(
-\beta\mu\right)  -1\right)  \right]  -\frac{C_{1}}{\beta\hbar\omega_{0}}%
\ln\left(  -\beta\mu\right)  \nonumber\\
&  +\frac{\Gamma\left(  1+s\right)  \left(  -\mu\right)  ^{s}}{\beta\left(
\hbar\omega_{y}\right)  ^{1+s}}\sum\nolimits^{\prime}\frac{1}{\left(  \kappa
n_{x}+n_{y}-\frac{\mu}{\hbar\omega_{y}}\right)  ^{1+s}}-\frac{C_{0}%
\Gamma\left(  s-1\right)  }{\left(  \beta\hbar\omega_{0}\right)  ^{2}}\left(
-\beta\mu\right)  -\frac{C_{1}\Gamma\left(  s\right)  }{\beta\hbar\omega_{0}%
}.\label{Nmu_HO}%
\end{align}
The sum of the spectrum is approximately%
\begin{equation}
\sum\nolimits^{\prime}\frac{1}{\left(  \kappa n_{x}+n_{y}-\frac{\mu}%
{\hbar\omega_{y}}\right)  ^{1+s}}\approx\sum\nolimits^{\prime}\left[  \frac
{1}{\left(  \kappa n_{x}+n_{y}\right)  ^{1+s}}-\frac{\left(  1+s\right)
\left(  -\mu\right)  }{\hbar\omega_{y}\left(  \kappa n_{x}+n_{y}\right)
^{2+s}}\right]  .
\end{equation}
The first term has been calculated in eq. (\ref{sum3}), and the second term is
also divergent at $s\rightarrow0$:
\begin{equation}
\frac{\Gamma\left(  1+s\right)  \left(  -\mu\right)  ^{s}}{\beta\left(
\hbar\omega_{y}\right)  ^{1+s}}\sum\nolimits^{\prime}\frac{\left(  1+s\right)
\left(  -\mu\right)  }{\hbar\omega_{y}\left(  \kappa n_{x}+n_{y}\right)
^{2+s}}=\frac{-\mu}{\kappa\beta\left(  \hbar\omega_{y}\right)  ^{2}}\left(
\frac{1}{s}+\ln\frac{-\mu}{\hbar\omega_{y}}-\gamma+\Delta_{2}\right)
,\label{sum4}%
\end{equation}
where
\begin{equation}
\Delta_{2}=\gamma+1+\frac{\pi^{2}}{6}\kappa-\sum_{k=1}^{\kappa-1}\frac
{k}{\kappa^{2}}\zeta\left(  2,\frac{k}{\kappa}\right)
\end{equation}
is only related to $\kappa$. However, the term with $\Gamma\left(  s-1\right)
$ in eq. (\ref{Nmu_HO}) is proportional to $\mu$ and should be included in
this approximation. Easy to check that the divergent term coming from the
gamma function and that in eq. (\ref{sum4}) are exactly canceled. Therefore
all the divergent terms of $s$ are canceled in eq. (\ref{Nmu_HO}):%
\begin{equation}
N=\frac{1}{-\beta\mu}+\frac{\zeta\left(  2\right)  }{\kappa\left(  \beta
\hbar\omega_{y}\right)  ^{2}}+\frac{1}{\beta\hbar\omega_{y}}\left[  \frac
{1}{2}\left(  1+\frac{1}{\kappa}\right)  \ln\frac{1}{\beta\hbar\omega_{y}%
}-\Delta\right]  -\frac{-\beta\mu}{\kappa\left(  \beta\hbar\omega_{y}\right)
^{2}}\left(  \ln\frac{1}{\beta\hbar\omega_{y}}+\Delta_{2}\right)
.\label{Nmu2_HO}%
\end{equation}
By using eqs. (\ref{N00_HO}) and (\ref{T0_HO}), it can be rewritten as%
\begin{equation}
N_{0}^{\left(  0\right)  }=\frac{1}{-\beta\mu}-\frac{6N}{2\pi^{2}}\left(
\frac{T}{T_{0}}\right)  ^{2}\left\{  \ln\left[  \frac{6\kappa N}{\pi^{2}%
}\left(  \frac{T}{T_{0}}\right)  ^{2}\right]  +2\Delta_{2}\right\}  \left(
-\beta\mu\right)  .
\end{equation}
Neglecting the higher-order terms, we solve the chemical potential as%
\begin{equation}
\mu\approx\mu_{c}\left[  \sqrt{1+\frac{\pi^{2}}{12}\left(  \frac{T_{0}}%
{T}\right)  ^{2}\frac{N_{0}^{\left(  0\right)  2}}{N\ln N}}-\frac{\pi}%
{2\sqrt{3}}\frac{T_{0}}{T}\frac{N_{0}^{\left(  0\right)  }}{\sqrt{N\ln N}%
}\right]  ,\label{mu_HO}%
\end{equation}
where%
\begin{equation}
\mu_{c}=-\sqrt{\frac{2}{\ln N}}\hbar\omega_{0}%
\end{equation}
is the chemical potential at the transition point.

\begin{figure}[ptb]
\begin{center}
\includegraphics[height=3in]{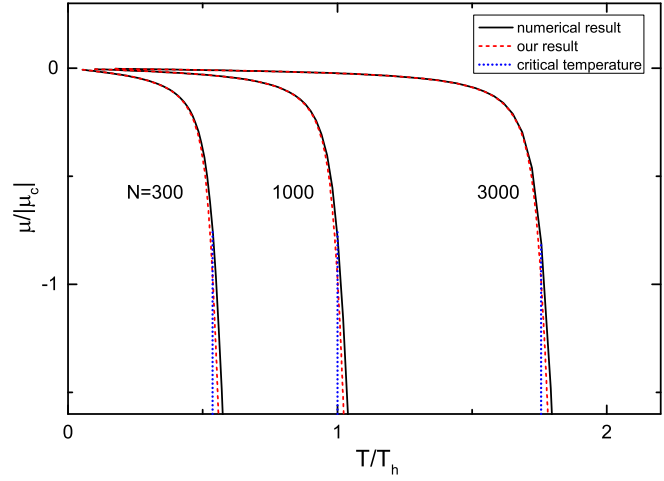}
\end{center}
\caption{The relation between chemical potential and temperature for different
$N$ in a two-dimensional isotropic harmonic trap ($\kappa=1 $). Our result
matches the numerical solution very well. The three sets of lines denote
$N=300,1000,3000$ from left to right.}%
\label{fig6}%
\end{figure}

\begin{figure}[ptb]
\begin{center}
\includegraphics[height=3in]{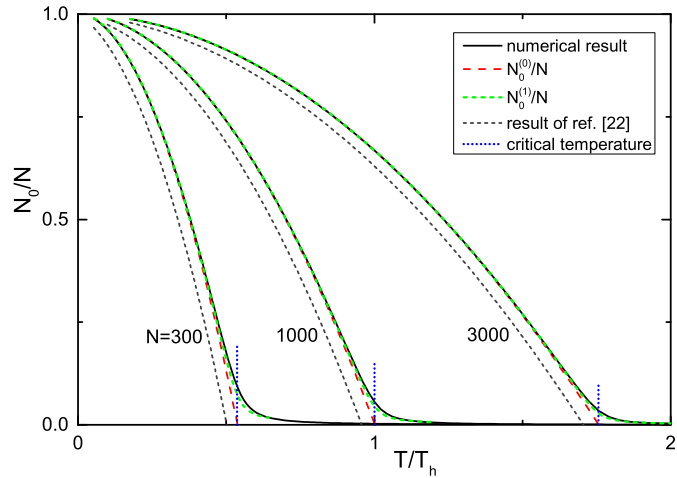}
\end{center}
\caption{The relation between condensate fraction and temperature for
different $N$ in a two-dimensional isotropic harmonic trap ($\kappa=1$). The
result of ref. \cite{Mullin}, the numerical solution, the zero- and
first-order approximations are plotted for $N=300,1000,3000$ from left to
right.}%
\label{fig7}%
\end{figure}

In fig. \ref{fig6} we plot the relation between the chemical potential and
temperature for different $N$. For $T<T_{c}$, eq. (\ref{mu_HO}) agrees with
the numerical solution quite good.

From eq. (\ref{mu_HO}), the first-order condensate fraction $N_{0}^{\left(
1\right)  }/N$ is straightforward according to eq. (\ref{N0}). In fig.
\ref{fig7} we plot the relation between the condensate fraction and
temperature for different $N$. At the critical temperature, the zero-order
condensate fraction vanishes, but the first-order one matches the numerical
solution very well.

\section{Comparison with the canonical ensemble\label{sec_CE}}

In the above sections, our discussion on BEC is in the grand canonical
ensemble. However, in a finite system, the fluctuation of particle number in
the grand canonical ensemble may be non-negligible. For investigating the
influence of fluctuation, we will consider the behavior of Bose gases in the
canonical ensemble and compare the result with the grand canonical ensemble.

There are many studies on the similarities and differences between different
ensembles for finite systems \cite{HHA,LGJJ,MF,GKP}. In this section, we will
take the two-dimensional harmonic trap as an example to show the difference
between these two ensembles.

In the canonical ensemble, the partition function of a $N$-particle system is%
\begin{equation}
Q\left(  N\right)  =\sum_{k}e^{-\beta E_{k}^{\left(  S\right)  }},
\label{PF_0}%
\end{equation}
where $E_{k}^{\left(  S\right)  }$ is the total energy of the $k$-th system in
the ensemble. However, the constraint of fixed particle number makes the exact
analytical form of partition function for a quantum system hard to obtain,
even for ideal gases. One method is to express the partition function by a
complex integral of the grand partition function as%
\begin{equation}
Q(N)=\frac{1}{2\pi i}%
{\displaystyle\oint}
z^{-N-1}\Xi(z)dz,
\end{equation}
where the integral path is a loop surrounding the original point. However,
although this integral can be approximately evaluated by the saddle point
method for large $N$, the exact integral can hardly be performed.

To give a direct comparison between different ensembles, we need the exact
partition function. For not very large $N$, this can be achieved by use of the
recursion relation \cite{LGJJ,MF}%
\begin{equation}
Q\left(  N\right)  =\frac{1}{N}\sum_{k=1}^{N}Q_{1}\left(  k\right)  Q\left(
N-k\right)  ,\text{ \ \ }\left(  Q\left(  0\right)  =1\right)  \label{PF}%
\end{equation}
where
\begin{equation}
Q_{1}\left(  k\right)  =\sum_{i}e^{-k\beta E_{i}}%
\end{equation}
is the partition function for a single particle at the temperature $T/k$.

We will take the condensate fraction as an example to compare with that in the
grand canonical ensemble. The average particle number in a state with energy
$E_{i}$ in the canonical ensemble can be expressed as \cite{LGJJ,MF}£¬%
\begin{equation}
\bar{N}_{i}=\frac{1}{Q\left(  N\right)  }\sum_{k=1}^{N}e^{-k\beta E_{i}%
}Q\left(  N-k\right)  .
\end{equation}
Combined with eq. (\ref{PF}), it will give the particle number in the ground
state and the condensate fraction.

\begin{figure}[ptb]
\begin{center}
\includegraphics[height=3in]{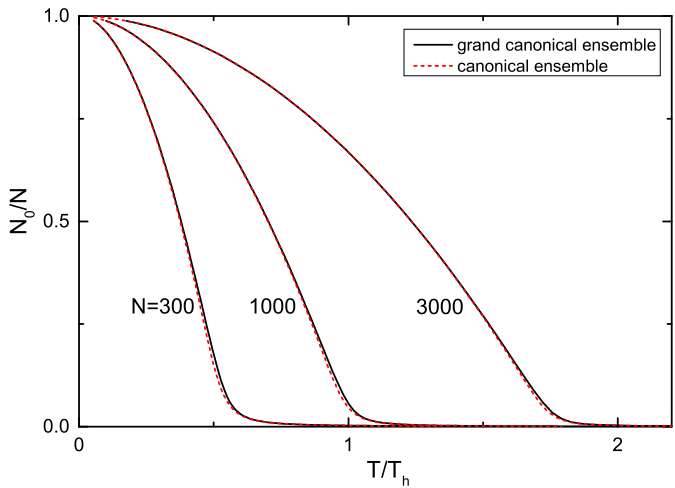}
\end{center}
\caption{The relation between condensate fraction and temperature in a
two-dimensional isotropic harmonic trap ($\kappa=1$) in different ensembles.
The exact numerical solutions in the grand canonical ensemble and the
canonical ensemble are plotted for $N=300,1000,3000$ from left to right.}%
\label{fig10}%
\end{figure}

In fig. \ref{fig10} we plot the numerical solutions of condensate fraction in
the grand canonical and canonical ensembles for different $N$ in a
two-dimensional harmonic trap. It is clear that for $N\sim10^{3}$, the
difference between these two ensembles is very small.

\section{Conclusion and discussion\label{sec4}}

In the above, by using an analytical continuation method to solve the
divergence problem in BEC, we discuss the low-temperature behavior of ideal
Bose gases in the two-dimensional anisotropic box and harmonic trap,
respectively. We show that the influence of boundaries and external potentials
can be dealt with by a general treatment. We obtain the critical temperature,
the condensate fraction and the chemical potential for Bose gases in these two
kinds of traps, respectively. The results are consistent with or better than
the corresponding studies in the literature, and they agree with the numerical
calculation well. To check the influence of fluctuation in the canonical
ensemble, we compare the condensate fraction in the grand canonical and
canonical ensembles. The result shows that for about $N\sim10^{3}$, the
difference between these two ensembles is negligible.

Although some previous studies also discussed the corrections to critical
temperature and condensate fraction in finite systems, our method is not an
order-of-magnitude estimate, so we can obtain more precise results, including
not only the leading correction. Besides, our method provides a general
treatment to the problem of BEC in finite systems. As long as the heat kernel
expansion is known, the critical temperature and the thermodynamic quantities
of the Bose gas can be calculated.

The grand potential of a finite system also contains divergent terms at
$\mu\rightarrow0$, and this problem can also be solved by similar treatment.
The analytical expressions of the grand potential and other thermodynamic
quantities below the transition point can be obtained as well. However, the
divergence problem is often not serious in the grand potential. For the two
cases considered in this paper, the divergence appears from the third term of
the grand potential. Therefore, our method will give the corrections to the
third terms. Such corrections are usually negligible, so their expressions are
not presented in this paper.

The advantage of our method is to remove the divergence at the transition
point, so the magnitude of the correction tightly depends on the specific
nature of the systems. For example, for the critical temperature, it gives the
second-order correction in the case of three-dimensional harmonic traps, which
is usually negligible \cite{Xie}. In a two-dimensional harmonic trap, the
correction is first-order and is expected to be observed in experiments. In a
two-dimensional box, since no phase transition exists in the thermodynamic
limit, the correction is zero-order and its influence is significant.

Recently, many experimental studies on BEC are performed in two-dimensional
traps, especially the BEC of quasiparticles, such as excitons in graphene and
surface exciton-polaritons. We hope that more precise experiments at this
field will test our results.

\bigskip

The author is very indebted to Prof. Wu-Sheng Dai for his help.
The author is grateful to an anonymous referee for helpful
comments and suggestions which greatly improved this paper. This
work is supported in part by NSF of China, under Project No.
11575125.

\section*{Appendix A: Asymptotic expansion of the Epstein $\zeta$-function
$E_{2}\left(  \sigma;a_{1},a_{2}\right)  $\label{app}}

According to ref. \cite{ER}, the Epstein $\zeta$-function%
\begin{align}
E_{2}\left(  \sigma;a_{1},a_{2}\right)   &  =-\frac{1}{2}a_{2}^{-\sigma}%
\zeta\left(  2\sigma\right)  +\frac{1}{2}a_{2}^{-\sigma}\sqrt{\frac{\pi a_{2}%
}{a_{1}}}\frac{\Gamma\left(  \sigma-1/2\right)  }{\Gamma\left(  \sigma\right)
}\zeta\left(  2\sigma-1\right) \nonumber\\
&  +\frac{2\pi^{\sigma}}{\Gamma\left(  \sigma\right)  }a_{1}^{-\frac{\sigma
}{2}-\frac{1}{4}}a_{2}^{-\frac{\sigma}{2}+\frac{1}{4}}\sum_{n_{1},n_{2}%
=1}^{\infty}n_{1}^{\sigma-\frac{1}{2}}n_{2}^{-\sigma+\frac{1}{2}}%
K_{\sigma-1/2}\left(  2\pi\sqrt{\frac{a_{2}}{a_{1}}}n_{1}n_{2}\right)
\label{Epstein}%
\end{align}
has a singularity $\sigma=1$, so we need the asymptotic expansion of Epstein
$\zeta$-function around $\sigma=1$.

Around $\sigma=1$, only the second term in eq. (\ref{Epstein}) is divergent,
which is%
\begin{align}
&  \frac{1}{2}a_{2}^{-\sigma}\sqrt{\frac{\pi a_{2}}{a_{1}}}\frac{\Gamma\left(
\sigma-1/2\right)  }{\Gamma\left(  \sigma\right)  }\zeta\left(  2\sigma
-1\right) \nonumber\\
&  \approx\frac{\pi}{4}\sqrt{\frac{1}{a_{1}a_{2}}}\left[  \frac{1}{\sigma
-1}+3\gamma+\psi\left(  \frac{1}{2}\right)  -\ln a_{2}\right]  .
\end{align}
The first and third terms in eq. (\ref{Epstein}) is convergent, so
substituting $\sigma=1$ into them gives%
\begin{equation}
-\frac{1}{2}a_{2}^{-1}\zeta\left(  2\right)  =-\frac{\pi^{2}}{12a_{2}}%
\end{equation}
and%
\begin{align}
&  2\pi a_{1}^{-\frac{3}{4}}a_{2}^{-\frac{1}{4}}\sum_{n_{1},n_{2}=1}^{\infty
}n_{1}^{\frac{1}{2}}n_{2}^{-\frac{1}{2}}K_{1/2}\left(  2\pi\sqrt{\frac{a_{2}%
}{a_{1}}}n_{1}n_{2}\right) \nonumber\\
&  =2\pi a_{1}^{-\frac{3}{4}}a_{2}^{-\frac{1}{4}}\frac{1}{2}\left(
\frac{a_{1}}{a_{2}}\right)  ^{1/4}\sum_{n_{1},n_{2}=1}^{\infty}\frac{1}{n_{2}%
}e^{-2\pi\sqrt{\frac{a_{2}}{a_{1}}}n_{1}n_{2}}\nonumber\\
&  =-\frac{\pi^{2}}{12a_{1}}-\frac{\pi}{\sqrt{a_{1}a_{2}}}\ln\eta\left(
i\sqrt{\frac{a_{2}}{a_{1}}}\right)  .
\end{align}
Therefore, around $\sigma=1$, we have%
\begin{align}
E_{2}\left(  \sigma;a_{1},a_{2}\right)   &  \approx\frac{\pi}{4\sqrt
{a_{1}a_{2}}}\frac{1}{\sigma-1}-\frac{\pi^{2}}{12}\left(  \frac{1}{a_{1}%
}+\frac{1}{a_{2}}\right) \nonumber\\
&  +\frac{\pi}{4\sqrt{a_{1}a_{2}}}\left[  3\gamma+\psi\left(  \frac{1}%
{2}\right)  -\ln\left(  a_{2}\eta^{4}\left(  i\sqrt{\frac{a_{2}}{a_{1}}%
}\right)  \right)  \right]  . \label{E.020}%
\end{align}

\end{document}